\documentclass[]{seismica}
\usepackage{todonotes}
\usepackage{bm}
\usepackage{amsfonts}

\title{Bayesian eikonal tomography using Gaussian processes}
\shorttitle{Bayesian eikonal tomography using Gaussian processes} 

\author[1]{Jack B. Muir
	\orcid{0000-0003-2617-3420}
	\thanks{Corresponding author: jack.muir@earth.ox.ac.uk}
}

\affil[1]{Department of Earth Sciences, University of Oxford}

\credit{Conceptualization}{Jack Muir}
\credit{Software}{Jack Muir}
\credit{Formal Analysis}{Jack Muir}
\credit{Visualization}{Jack Muir}
\credit{Writing — Original draft}{Jack Muir}

\nolinenumbers
\begin{document}

\makeseistitle{
    \begin{summary}{Abstract}
        Eikonal tomography has become a popular methodology for deriving phase velocity maps from surface wave phase delay measurements. Its high efficiency makes it popular for handling datasets deriving from large-N arrays, in particular in the ambient-noise tomography setting. However, the results of eikonal tomography are crucially dependent on the way in which phase delay measurements are predicted from data, a point which has not been thoroughly investigated. In this work, I provide a rigorous formulation for eikonal tomography using Gaussian processes (GPs) to smooth observed phase delay measurements, including uncertainties. GPs allow the posterior phase delay gradient to be analytically derived. From the phase delay gradient, an excellent approximate solution for phase velocities can be obtained using the saddlepoint method. The result is a fully Bayesian result for phase velocities of surface waves, incorporating the nonlinear wavefront bending inherent in eikonal tomography, with no sampling required. The results of this analysis imply that the uncertainties reported for eikonal tomography are often underestimated. 
    \end{summary}
    \begin{summary}{Non-technical summary}
        Eikonal tomography is an imaging method that uses slight variations between seismic waves trapped at the surface of the Earth to infer information about the properties beneath the surface. To be able to perform the best possible eikonal tomography, we need to be able to predict in between measurements of these variations at different seismic recording stations as best we can. Furthermore, end-users of seismic tomography require information about the uncertainty of the images. In this paper, I perform this prediction using Gaussian processes (GPs), a method with particularly nice mathematical properties. The GP prediction results in robust uncertainty measurements for our imaging problem without many of the computational difficulties associated with other uncertainty quantification methods. 
    \end{summary}
    }

\section{Introduction}

Surface wave tomography is a cornerstone imaging technique for the investigation of the crust and upper mantle. However, due to the significant non-planarity of scattered surface waves, interpretation of surface wave data is not straightforward \cite[e.g.,][]{wielandtPropagationStructuralInterpretation1993a}. Despite this issue, the increasing proliferation of dense seismic arrays, combined with the advent of ambient-noise correlation methods, has motivated intense study into surface wave tomographic techniques. To ameliorate the great cost of nonlinear ray tracing for large inverse problems, a large part of this study has focused on methods that derive surface wave properties from only local information contained in the wavefield. Beginning with a wavefield perturbation approach \cite[e.g.,][]{friederichNonPlaneGeometriesSeismic1994, friederichInterpretationSeismicSurface1995, pollitzObservationsInterpretationFundamental2008}, theoretical efforts in local surface wave inversion have since concentrated on direct measurement of wavefield derivatives \cite[e.g][]{linEikonalTomographySurface2009, linHelmholtzSurfaceWave2011, deridderSurfaceScholteWave2015, deridderFullWaveField2018}. Likely owing to its simplicity, the most popular extant method is eikonal tomography \cite[]{linEikonalTomographySurface2009}, which relies on the determination of the wavefield phase gradient across an entire local or regional array. For a single surface wave mode propagating with phase velocity $C_p$, frequency $\omega$, phase delay $\tau$ and amplitude $A$, the Helmholtz equation implies that \cite[]{trompVariationalPrinciplesSurface1993}
\begin{equation}
\frac{1}{C_p^2} = |\nabla \tau|^2 - \frac{\nabla^2 A}{\omega^2 A}.
\label{eq:helmholtz}
\end{equation}
Simplifying this relationship under the assumption that the frequency of the wave is large compared to perturbations in the wave amplitude gives us the eikonal equation: 
    \begin{equation}
    C_p = \frac{1}{|\nabla \tau|}. 
    \label{eq:eikonal}
    \end{equation}
Eikonal tomography uses Equation \ref{eq:eikonal} to directly infer local phase velocity from local phase gradient. A distinction compared to local gradiometry is that calculation of the phase gradient is performed simultaneously for all desired locations by fitting a delay curve across an array, rather than by local analysis of sub-arrays \cite[e.g.,]{langstonSpatialGradientAnalysis2007}. The assumption that the wavefront is smooth relative to frequency is strong, but the difficulty associated with measuring wavefront curvature accurately has ensured that eikonal tomography remains a central technique in array analysis. Application of eikonal tomography in practice has typically resulted in images comparable to other tomographic methods and Helmholtz tomography (which uses Equation \ref{eq:helmholtz} directly), especially when results are averaged azimuthally \cite[]{bodinResolutionPotentialSurface2008, linEikonalTomographySurface2009, lehujeurValidityEikonalEquation2020}. 

In this work, I employ Gaussian process theory \cite[]{rasmussenGaussianProcessesMachine2006} to derive semi-analytic closed-form approximations for the posterior distribution of eikonal-equation-based phase velocity measurements using the saddlepoint method \cite[]{butlerSaddlepointApproximationsApplications2007}. In this case, semi-analytic means that the posterior approximations have a single parameter that must be solved using constrained minimization techniques --- no Monte Carlo methods need be used. As a result, the approximate posterior can be calculated very quickly. As an intermediate result, I derive fully analytic posteriors for the gradient of phase delay. The delay gradient posteriors can be sampled using standard multivariate normal random number generators, which provides an efficient way to compute arbitrary statistics of the GP posterior when the semi-analytic approximations are difficult to obtain. 

\section{Eikonal tomography from derivatives of Gaussian processes}

The least well-defined problem in eikonal tomography is how to go from point measurements of phase delay to the phase delay gradient map \cite[]{linEikonalTomographySurface2009}. It is in this process that the practitioner has the greatest control over the resulting phase velocity map; intuitively, we can immediately see that over-smoothing the map will result in a measurement of $C_p$ that is too large; conversely, maps that are too rough will result in too small $C_p$. Past studies have typically employed splines (either in tension \cite[e.g.,][]{linEikonalTomographySurface2009, linHelmholtzSurfaceWave2011} or smoothing \cite{chevrotEikonalSurfaceWave2022}) to perform prediction. The spline framework is a robust general interpolation or smoothing method, however in its basic formulation it gives a single maximum-likelihood estimate of the prediction, with no associated uncertainty information. 

This study is aims to place the problem of estimating an optimal phase gradient map on a robust Bayesian footing, where all assumptions are explicit, adjustable, and optimizable in the face of the data. In this study, the problem of predicting phase delay measurements is posed as a Gaussian process (GP) regression --- we will see that this framework meets the desiderata for estimating phase gradients. GPs are a particular framework for defining distributions over function spaces \cite[]{rasmussenGaussianProcessesMachine2006}. GPs have the property that any finite collection of points sampled from them will have a multivariate Gaussian distribution. A GP is defined by a mean function $f(x)$ and covariance function $k(x,x')$, which generate the mean and covariance matrix of a finite collection of points drawn from the GP. In the context of regression, this leads to a powerful result --- if we assume a GP prior for an unknown function, and we then observe data with a Gaussian likelihood, the posterior distribution for the unknown function will also be a GP. Thus, GPs fully generalize finite linear regression and Gaussian inverse problems to the function space setting \cite[]{valentineGaussianProcessModels2020,valentineGaussianProcessModels2020a}. As differentiation is a linear operation, derivatives of GPs are again also GPs. We will use these properties to derive closed-form posterior distributions for the derivatives of observed data under a GP prior. While the motivating example is eikonal tomography, these techniques are applicable to regression problems generally. Derivatives of GPs have long been used in the dynamical control community \cite[e.g.][]{solakDerivativeObservationsGaussian2002, rasmussenGaussianProcessesSpeed2003} Closer in spirit to seismology, GP derivatives have also been applied to the identification of geodetic transients \cite[]{hinesRevealingTransientStrain2018}. The presentation described here is generalized from \cite{mchutchonNonlinearModellingControl2014}.

\begin{figure}
    \centerline{\includegraphics{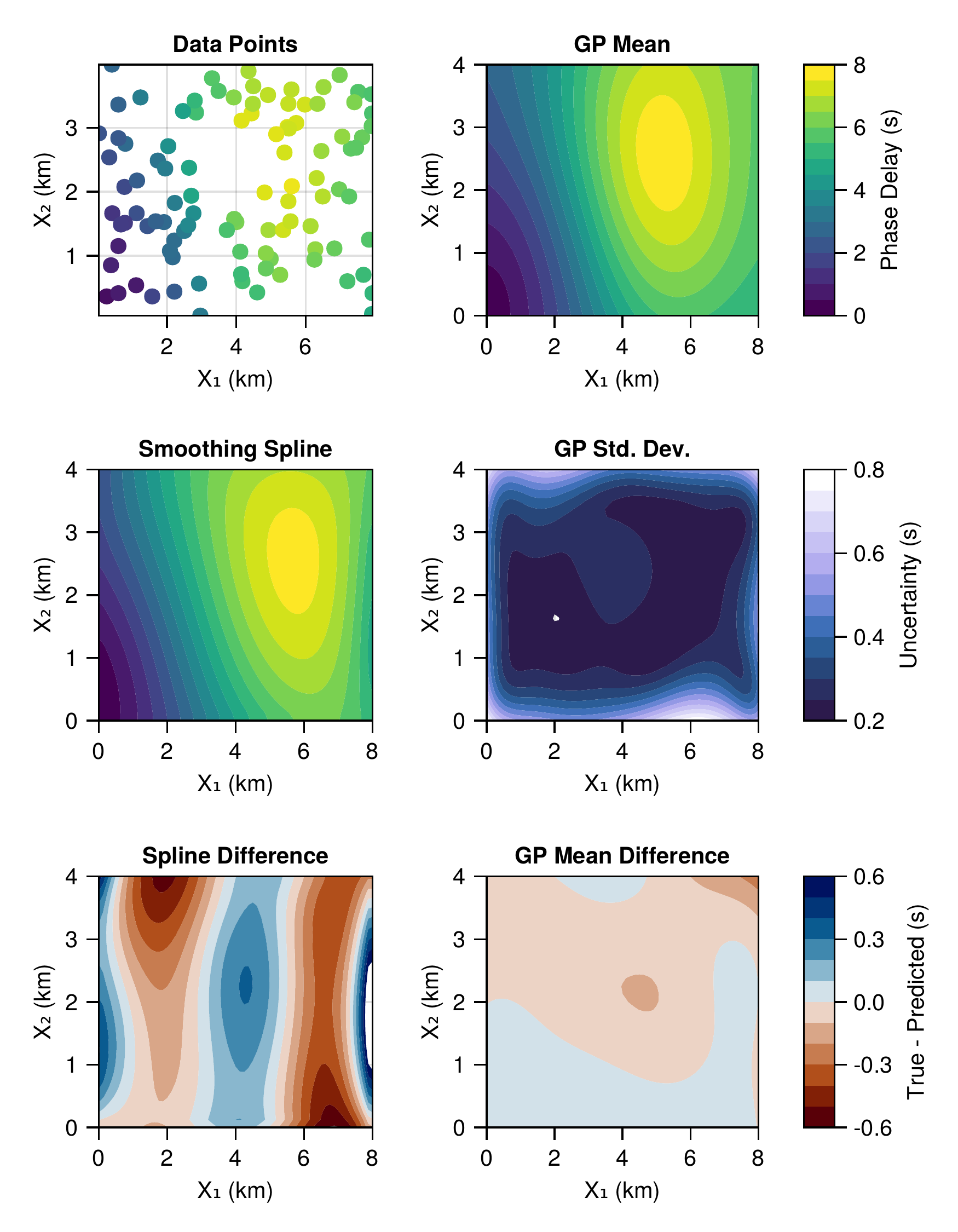}}
    \caption{Comparison of the GP posterior (showing mean and point-wise standard deviation) of the phase delay with a smoothing-spline based solution for an example phase delay data set with 100 randomly distributed points and 0.2 s Gaussian noise. There are notable differences in the estimated phase delay, especially where there are gaps in the data coverage. The colouring of the difference plots is arranged according to the usual seismic convention of blue being a fast and red being slow; in this case blue means that the predicted arrival is fast compared to the truth and vice versa.}
    \label{fig:gp_spline}
\end{figure}

\begin{figure}
    \centerline{\includegraphics{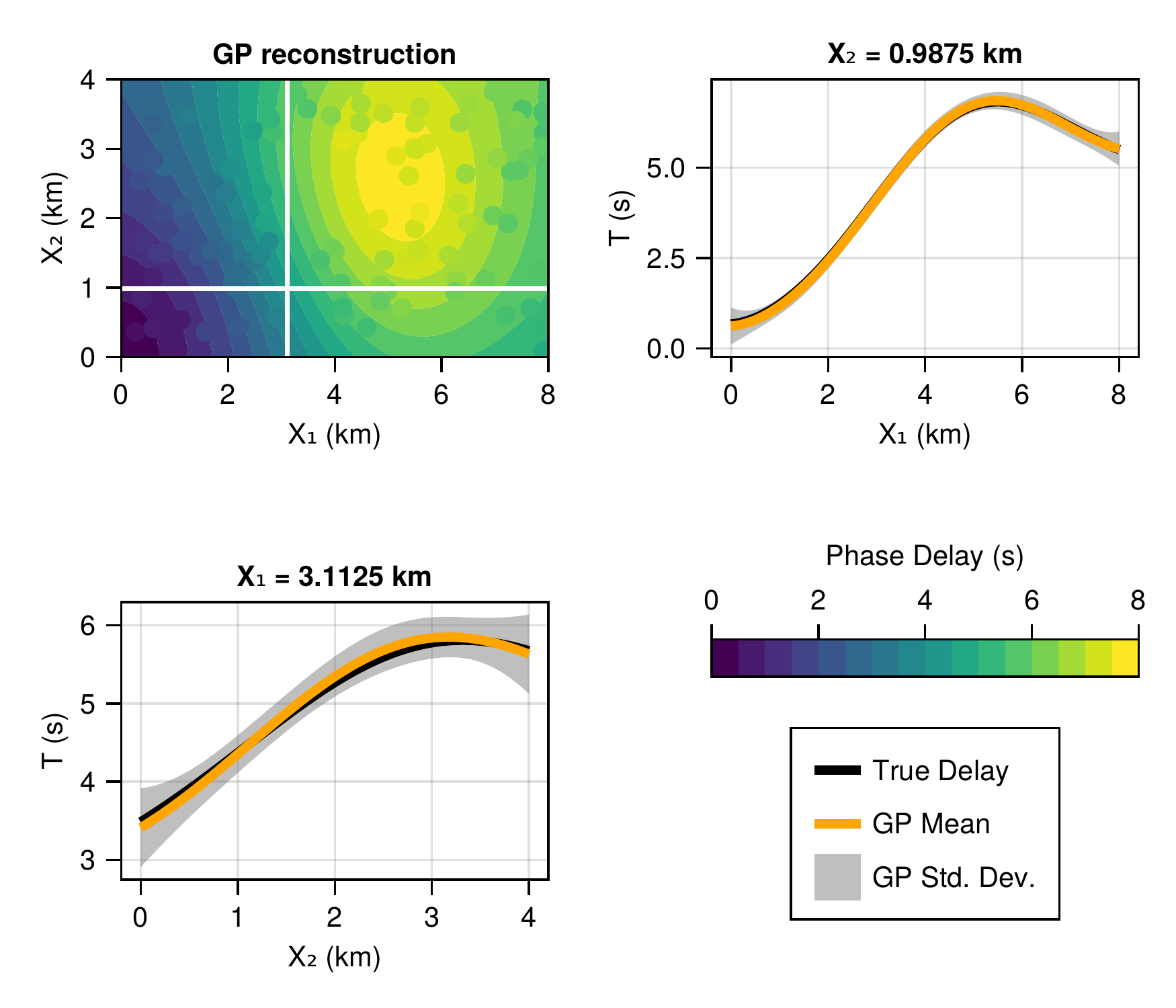}}
    \caption{Cross-sections through the GP reconstruction, showing the true phase delay (black), GP mean (orange) and standard deviation (grey). The GP reconstruction is overlaid with the noisy observed delay values. The GP posterior closely follows the true phase delay curve, with substantially higher uncertainty near the edges of the domain, even before extrapolation.}
    \label{fig:gp_rec}
\end{figure}

In general, bold font refers to 1D collections of data and capitals to matrices. Boldfont capitals are therefore collections of $n$ data in $d$ coordinates and will have dimensions $n\times d$. Coordinates (i.e., $x$) may be vector quantities but will not be boldfont. To begin, assume that there are measurements $(\bm{X}, \bf{y})$ of the observed phase delay $\bf{y}$ at points $\bm{X}$. Assume that the data $\bf{y}$ are noisy; for the purposes of exposition this is taken to be identically distributed Gaussian noise $\eta$ with the distribution $N(0, \sigma)$, but arbitrary multivariate Gaussian noise distributions are also easily handled by GP theory. This implies that there is an unknown true phase delay field $\tau(x)$ with 
    \begin{equation}
        \bf{y} = \tau(\bf{X}) + \eta.
    \end{equation}
The objective of eikonal tomography is to know the field $\tau(x)$ so that we can differentiate it and get $C_p$. I assume that 
    \begin{equation}
    \tau(x) = \tau_0(x) + f(x)
    \end{equation}
where $f$ is a zero-mean GP and $\tau_0(x)$ is a reference phase delay field, for example for a laterally homogeneous medium. Therefore, $\tau(x)$ is a GP with mean $\tau_0(x)$. 
    \begin{equation}
    \tau(x) \sim GP(\tau_0(x), k(x, x')), 
    \end{equation}
where $k(x, x')$ is the assumed covariance function. For the examples in this work, I will use a squared-exponential kernel with independent length scales in each dimension for the covariance function:
\begin{equation}
    k(x,x') = \rho^2 \exp\left(-\sum_{i=1}^d \frac{(x_i - x_i')^2}{2 l_i^2}\right).
\end{equation}
This covariance function promotes very smooth fields (it is infinitely differentiable), and provides a degree of flexibility due to the independent length scales. I also assume that $\tau_0(x) = s_0|x|$ for a fixed reference slowness $s_0$. Let $K_{\bm{X}\bm{X}'}$ be the matrix of evaluating $k$ with rows given by $\bm{X}$ and columns by $\bm{X}'$. The fundamental idea of GP regression is that, given this problem setup, then the observed data $\bm{y}$ and the predicted data $\tau(\bm{X}')$ has the joint multivariate Gaussian distribution
    \begin{equation}
    \begin{bmatrix}\bf{y} \\ \tau(\bm{X}') 
    \end{bmatrix} \sim N\left(\begin{bmatrix}\tau_0(\bm{X}) \\ \tau_0(\bm{X}') 
    \end{bmatrix}, 
    \begin{bmatrix}K_{\bm{X}\bm{X}} + \sigma^2I & K_{\bm{X}\bm{X}'} \\ K_{\bm{X}'\bm{X}} & K_{\bm{X}'\bm{X}'}
    \end{bmatrix}.
    \right)
    \end{equation}
By conditioning $\tau(\bm{X}')$ on the observed data $\bm{y}$ we have \cite[]{rasmussenGaussianProcessesMachine2006}
    \begin{equation}
        \tau(\bm{X}') | \bm{y} \sim N(\tau_0(\bm{X}') + K_{\bm{X}'\bm{X}}(K_{\bm{X}\bm{X}} + \sigma^2I)^{-1}(\bm{y}-\tau_0(\bm{X})), K_{\bm{X}'\bm{X}'}-K_{\bm{X}'\bm{X}}(K_{\bm{X}\bm{X}} + \sigma^2I)^{-1}K_{\bm{X}\bm{X}'}).
    \end{equation}
Note that data error models with Gaussian covariance just require replacing $\sigma^2I$ with $C_D$. Figure \ref{fig:gp_spline} shows an example application of GP regression for obtaining $\tau(x) | \bf(y)$, with comparison to the  approach based on regression using  splines \cite[e.g.,][]{linEikonalTomographySurface2009, linHelmholtzSurfaceWave2011} --- in this case, using smoothing splines \cite[e.g.,][]{chevrotEikonalSurfaceWave2022}. This example emulates a typical local surface wave application, using 100 data points uniformly distributed within the inversion region with 0.2 s added Gaussian noise. The synthetic phase delay field is strongly perturbed away from the reference model to highlight differences between the two methods. The GP mean and standard deviation are given analytically, and show substantial differences with the smoothing spline fit --- here, the spline smoothing parameter is automatically set by the FitPack routine \cite[]{dierckxCurveSurfaceFitting1995}. In comparison to the GP, the spline performs less well, especially in areas of data gaps. Figure \ref{fig:gp_rec} compares the GP reconstruction with the true values of the phase delay map. T The GP mean closely fits the true values, although the level of uncertainty becomes quite substantial near the edges of the domain. 

\begin{figure}
    \centerline{\includegraphics{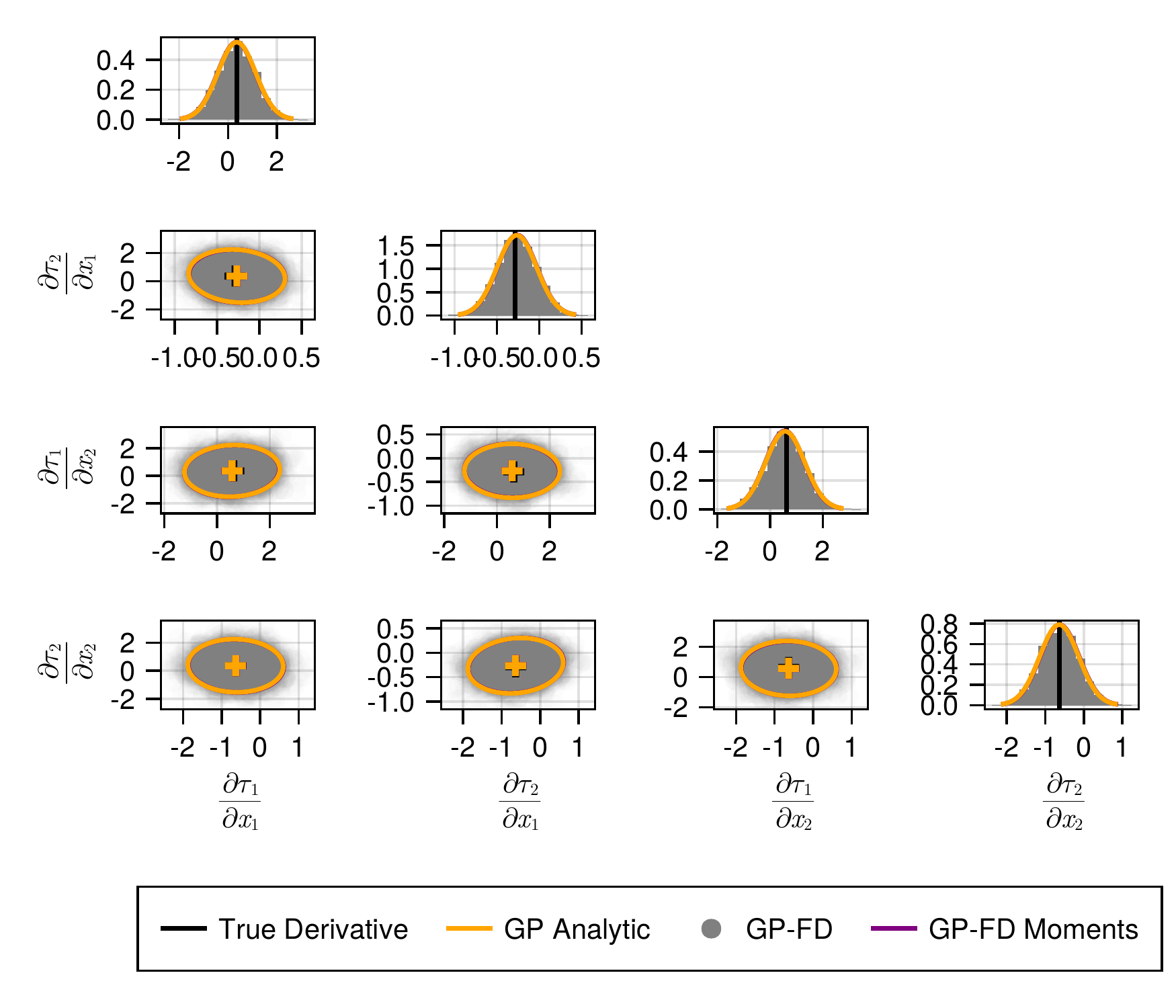}}
    \caption{Corner plot showing the covariance of derivatives at two test points, and their individual histograms. The test points are $\tau_1$ at (0.1875,0.3875), and $\tau_2$ at (5.5, 3.5). Black crosses and lines show the true value of the derivatives. Orange lines show the analytical GP based solutions derived in this paper, with ellipses drawn at the 95\% credible level and crosses showing the mean. Grey circles and histograms show finite-difference (FD) based derivatives using Monte-Carlo samples of the GP posterior for phase delay, and red crosses and ellipses show the mean and estimated covariance at 95\% confidence from the FD draws.}
    \label{fig:gp_derv}
\end{figure}

\begin{figure}
    \centerline{\includegraphics{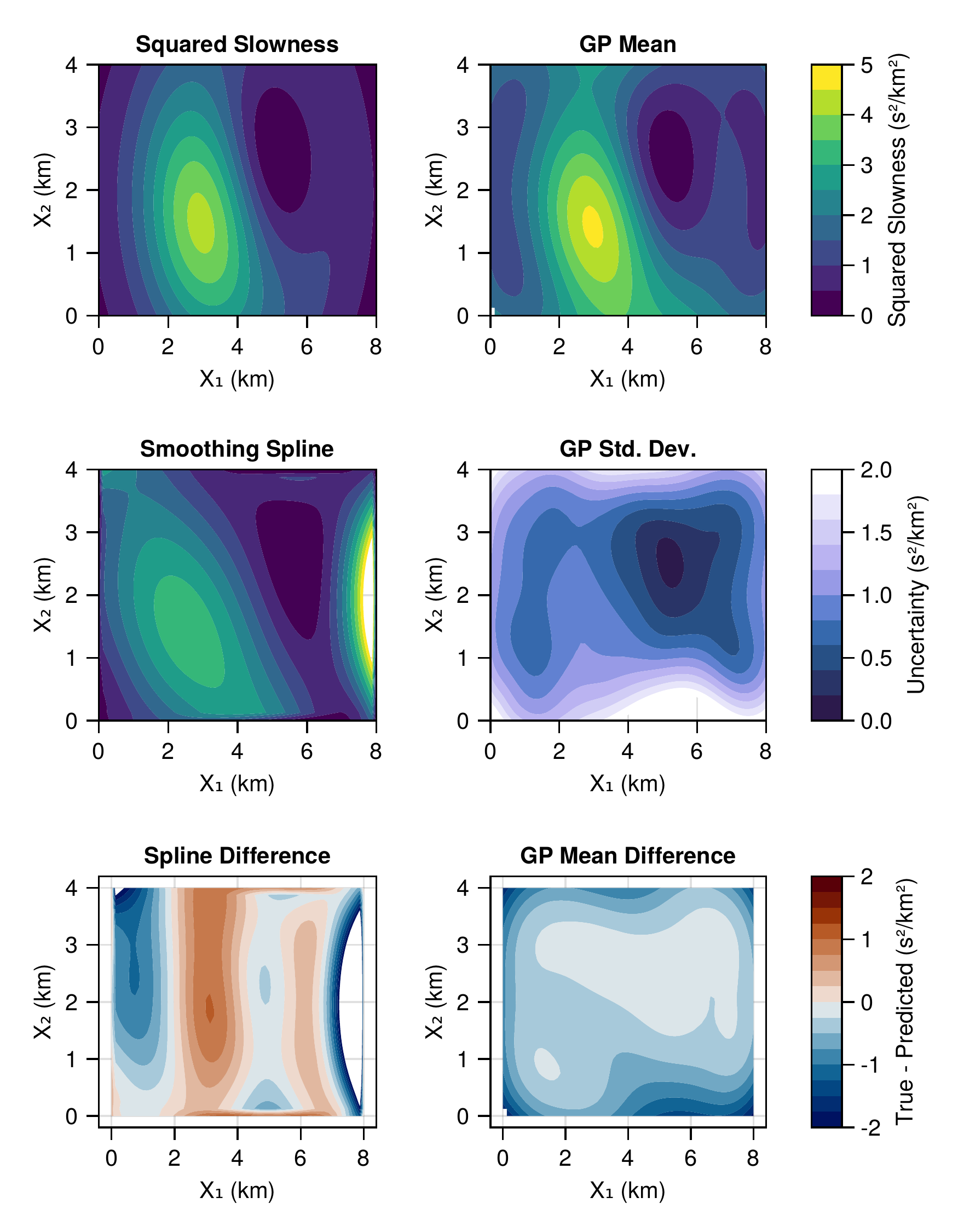}}
    \caption{Comparison of the true squared slowness against results calculated using a squared-exponential Gaussian process with tuned hyperparameters. The GP mean and standard deviation are calculated by drawing 100,000 predicted travel time gradients. The spline squared slowness has been calculated using 5$^{th}$ order centred finite differences. The GP result has a mean closer to the truth, and additionally adds uncertainty information, when compared to the smoothing spline. The colouring of the difference plots is arranged according to the usual seismic convention of blue being a fast and red being slow; in this case blue means that the predicted slowness is smaller compared to the truth and vice versa; note that this induces a colour flip compared to Figure \ref{fig:gp_spline}.}
    \label{fig:gp_k2}
\end{figure}

I  can now calculate expectation values for the derivatives; note that from now on I implicitly condition on $\bm{y}$ but will not write it out for ease of notation, unless it seems particularly germane to do so. Since differentiation is a linear operation, and linear operations acting on  normal distributions result in normal distributions, the components of $\nabla \tau$ must also be normally distributed, and are completely specified by their mean and covariance. The collection of means for component $i$ are immediately given by recognizing that as the expectation operator is also linear, it commutes with the derivative operator:
    \begin{equation}
    \mathbb{E}\left[\frac{\partial \tau(\bm{X}')}{\partial x'_i}\right] = \frac{\partial  \mathbb{E}\left[\tau(\bm{X}')\right]}{\partial x'_i} = \frac{\partial\tau_0(\bm{X}')}{\partial x'_i} + \frac{\partial K_{\bm{X}'\bm{X}}}{\partial x'_i}(K_{\bm{X}\bm{X}} + \sigma^2I)^{-1}(\bm{y}-\tau_0(\bm{X})).
    \end{equation}
Note that the mean value of the derivatives are calculated independently for each dimension; however as we will see they do have covariance between output points and between dimensions. For the covariance, consider $n\times n$ blocks of the covariance matrix of size $nd\times nd$ where $d$ is the dimension and $n$ is the number of output points. Note that I choose to order the hierarchy of the covariance matrix first by derivative coordinate, and second by data point index, as it makes the notation more convenient. As the covariance is bilinear, 
    \begin{equation}
        \text{Cov}\left(\frac{\partial \tau(\bm{X}')}{\partial x'_i}, \frac{\partial \tau(\bm{X}'')}{\partial x''_j}\right) = \frac{\partial^2 \text{Cov}(\tau(\bm{X}'), \tau(\bm{X}''))}{\partial x'_i \partial x''_j}
        \label{eq:gp_d_mean}
    \end{equation}
where I introduce the dummy variable $x''$ to represent the second argument in the covariance ($\bm{X}' = \bm{X}''$, but we want to formally differentiate in respect to the second slot only when using $x''$). Continuing on, 
    \begin{align}
        \frac{\partial^2 \text{Cov}(\tau(\bm{X}'), \tau(\bm{X}''))}{\partial x'_i \partial x''_j} & = \frac{\partial^2 \left(K_{\bm{X}'\bm{X}''}-K_{\bm{X}'\bm{X}}(K_{\bm{X}\bm{X}} + \sigma^2I)^{-1}K_{\bm{X}\bm{X}''} \right) }{\partial x'_i \partial x''_j} \nonumber\\
        & = \frac{\partial^2 K_{\bm{X}'\bm{X}''}}{\partial x_i' \partial x_j''}-\frac{\partial K_{\bm{X}'\bm{X}}}{\partial x_i'}(K_{\bm{X}\bm{X}} + \sigma^2I)^{-1}\frac{\partial K_{\bm{X}\bm{X}''}}{\partial x_j''}. 
        \label{eq:gp_d_cov}
    \end{align}
So that I can compress the notation somewhat, let us define $\hat{K}_{\bm{X}\bm{X}} = K_{\bm{X}\bm{X}} + \sigma^2I$ and $\Delta\bm{y} = \bm{y}-\tau_0(\bm{X})$. For the 2D case investigated here (noting that higher dimensions immediately generalize), the conditional posterior is a multivariate Gaussian with mean given by Equation \ref{eq:gp_d_mean} and covariance given by Equation \ref{eq:gp_d_cov}:
    \begin{align}
    \nabla \tau(\bm{X}')| \bm{y} & = 
    \begin{bmatrix} \frac{\partial \tau(\bm{X}')}{\partial x_1} \\ \frac{\partial \tau(\bm{X}')}{\partial x_1} \end{bmatrix} | \bm{y} \nonumber \\ 
    &\sim N\left(\begin{bmatrix} 
        \frac{\partial\tau_0(\bm{X}')}{\partial x'_1} + \frac{\partial K_{\bm{X}'\bm{X}}}{\partial x'_1}\hat{K}_{\bm{X}\bm{X}}^{-1}\Delta\bm{y} \\ 
        \frac{\partial\tau_0(\bm{X}')}{\partial x'_2} + \frac{\partial K_{\bm{X}'\bm{X}}}{\partial x'_2}\hat{K}_{\bm{X}\bm{X}}^{-1}\Delta\bm{y}
    \end{bmatrix}, 
    \begin{bmatrix}
        \frac{\partial^2 K_{\bm{X}'\bm{X}''}}{\partial x_1' \partial x_1''}-\frac{\partial K_{\bm{X}'\bm{X}}}{\partial x_1'}\hat{K}_{\bm{X}\bm{X}}^{-1}\frac{\partial K_{\bm{X}\bm{X}''}}{\partial x_1''} & \frac{\partial^2 K_{\bm{X}'\bm{X}''}}{\partial x_1' \partial x_2''}-\frac{\partial K_{\bm{X}'\bm{X}}}{\partial x_1'}\hat{K}_{\bm{X}\bm{X}}^{-1}\frac{\partial K_{\bm{X}\bm{X}''}}{\partial x_2''} \\
        \frac{\partial^2 K_{\bm{X}'\bm{X}''}}{\partial x_2' \partial x_1''}-\frac{\partial K_{\bm{X}'\bm{X}}}{\partial x_2'}\hat{K}_{\bm{X}\bm{X}}^{-1}\frac{\partial K_{\bm{X}\bm{X}''}}{\partial x_1''} & 
        \frac{\partial^2 K_{\bm{X}'\bm{X}''}}{\partial x_2' \partial x_2''}-\frac{\partial K_{\bm{X}'\bm{X}}}{\partial x_2'}\hat{K}_{\bm{X}\bm{X}}^{-1}\frac{\partial K_{\bm{X}\bm{X}''}}{\partial x_2''}
    \end{bmatrix}\right),
    \label{eq:derivatives}
    \end{align}
which is an exact distribution for the derivatives evaluated at $\bm{X}'$. Figure \ref{fig:gp_derv} shows the mean and covariance structure for the derivatives at two test points calculated using the above theory, compared to the true derivative of the phase delay, and finite-difference estimates computed using random draws of the GP estimate of the phase delay (i.e., Monte-Carlo finite-difference derivatives). Both the analytic and Monte-Carlo results closely agree with each other and with the true values for the derivatives. In Figure \ref{fig:gp_k2}, I use the multivariate normal posterior for the derivatives to generate samples of the posterior for the squared slowness and compare it against the predictions from the smoothing spline. The GP posterior is in this case more accurate than the spline result, and also delivers uncertainty information. 

Unfortunately, it turns out that this is as far as it is possible to go with exact distributions, as the velocity is a nonlinear function of the gradients in eikonal tomography. Thankfully, however, there is well-developed theory for approximating quadratic forms of normal random variables, and as $\frac{1}{C_p^2} = (\nabla\tau )^2$, which is a quadratic form of a normal random variable, it may be possible to try for a good approximation to the velocity. Before deriving one, however, there are two important issues to investigate --- setting hyperparameters, and closed forms for the expectation value of velocity. 

\subsection{Finding good values for GP hyperparameters}

The hyperparameters of the GP may be optimized by maximizing the log marginal likelihood of observations, where the marginalization is performed over the unknown function values $\tau(\bm{X})$ (\cite{rasmussenGaussianProcessesMachine2006}). This gives the type-II maximum likelihood estimate; the hyperparameters have a point-estimate, whereas the function values have a full posterior distribution given that point-estimate. The log marginal likelihood for GP regression is given by
    \begin{equation}
    \log p(\bm{y}|\theta, \bm{X})  = -\frac{1}{2}\Delta \bm{y}^T \hat{K}_{\bm{X}\bm{X}}^{-1}(\theta) \Delta \bm{y} - \frac{1}{2}|\hat{K}_{\bm{X}\bm{X}}(\theta)| - \frac{n}{2}\log(2\pi), 
    \end{equation}
where the covariance matrix $\hat{K}_{\bm{X}\bm{X}}(\theta)$ is treated as a function of the hyperparameters $\theta$, and $n$ is the number of data. Intuitively, the log marginal likelihood parsimoniously balances data misfit (the first term) with the level of uncertainty (the second term). For a 2D squared-exponential kernel with independent length scales, independent Gaussian data noise, and a laterally homogeneous medium as a reference model, the hyperparameters are $\theta = (\rho, l_1, l_2, \sigma, s_0)$.  

\subsection{A special exact case for eikonal tomography: The expectation value of squared slowness given normally distributed derivatives}

Consider without loss of generality a 2D case. The squared slowness is given by $1/C_p^2 = \left(\frac{\partial \tau}{\partial x_1}\right)^2 + \left(\frac{\partial \tau}{\partial x_2}\right)^2 = \tau_{x_1}^2 + \tau_{x_2}^2$. Assume the phase gradient is given by a multivariate Gaussian random variable 
    \begin{equation}
    \tau_x =  \begin{bmatrix} \tau_{x_1} \\ \tau_{x_2} \end{bmatrix}\sim N\left(\begin{bmatrix}\mu_{x_1}\\ \mu_{x_2}\end{bmatrix}, \begin{bmatrix}\sigma^2_{x_1} & \nu_{x_1x_2}\\ \nu_{x_1x_2} & \sigma^2_{x_2}\end{bmatrix}\right) = N(\mu, \Sigma)
    \end{equation}
that describes the joint distribution of the two derivatives $\tau_{x_1}, \tau_{x_2}$, and let $S^2$ be the random variable describing the distribution of slowness squared. This is, for example, the distribution that arises for the derivatives of a single point conditioned on observations under GP regression as described above. Then $\mathbb{E}[S^2]  = \mathbb{E}[\tau_x^t \tau_x]$. Note that $Cov[\tau_x, \tau_x] = \mathbb{E}[\tau_x\tau_x^T] - \mathbb{E}[\tau_x]\mathbb{E}[\tau_x]^T$. As the slowness squared is a scalar, I can take the trace to proceed as follows, following \cite{kendrickStochasticControlEconomic2002}: 
    \begin{align}
        \mathbb{E}[S^2] & = \mathbb{E}[\tau_x^T \tau_x]\nonumber\\
        & = \mathbb{E}[tr(\tau_x^T \tau_x)]\nonumber\\
        & = tr(\mathbb{E}[\tau_x^T \tau_x]) \nonumber \\
        & = tr(\mathbb{E}[\tau_x]\mathbb{E}[\tau_x]^T+Cov[\tau_x, \tau_x]) \nonumber \\
        & = tr( \mu\mu^T+\Sigma) \nonumber \\
        & = \mu_{x_1}^2 + \mu_{x_2}^2 + \sigma_{x_1}^2 + \sigma_{x_2}^2 \\
        & > \mu_{x_1}^2 + \mu_{x_2}^2  = (\mathbb{E}[\tau_x])^2 + (\mathbb{E}[T_{x_2}])^2\nonumber
    \end{align}
It is instructive to note that the expectation value of squared slowness is strictly greater than the sum-of-squares of the mean derivatives, so that velocities are ``biased'' lower after accounting for errors. Note that this is true for any calculation that assumes the derivatives have a Gaussian distribution, not just the Gaussian process framework analysed here.

%
%

\section{Approximation of the posterior using the saddlepoint method}

\begin{figure}
    \centerline{\includegraphics{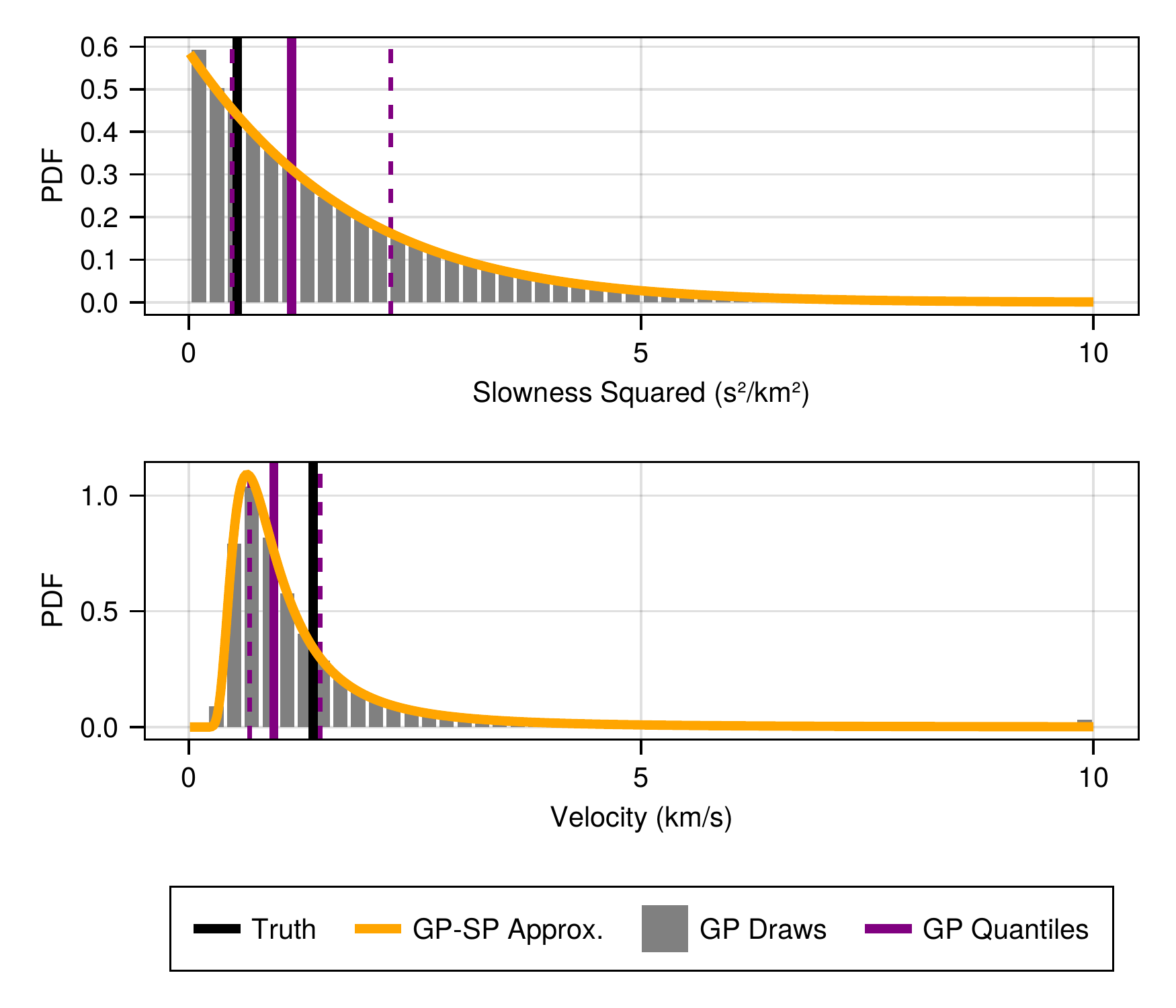}}
    \caption{Comparison of the empirical CDF and PDF (grey) for the squared slowness and phase velocity for the point at (0.1875,0.3875) with the saddlepoint (SP) approximation (orange). For the PDF, the true value is also shown in black and the median, 25$^{th}$ and 75$^{th}$ percentiles of the empirical PDF are shown in purple. The empirical distributions are truncated between 0.01 and 10 for plotting purposes.}
    \label{fig:gp_sp}
\end{figure}

The analytic results obtained for the derivative $\nabla \tau$ have already given us a great deal. Any expectation value that depends on these derivatives (in particular, moments of the phase velocity) can be calculated using the Monte-Carlo method --- i.e., by drawing many random samples of $\nabla \tau$ and then calculating the desired statistics on this random sample. Because it is possible to draw directly from the posterior of $\nabla \tau$ given Equation \ref{eq:derivatives}, every sample can be used and is independent (unlike in Markov-Chain Monte-Carlo). As such, these expectation values will usually converge quickly. However, there are cases where it is still useful to have approximations of the posterior that can be even more quickly calculated; for instance if the eikonal tomography derived phase velocities are being used in a joint inverse problem, or if accurate statistics for extreme values need to be calculated. A frequently used simple approximation would be to us Laplace's method directly on the posterior distribution for $||\nabla \tau||^2$ or $C_p$. The approximate posterior under using this technique is the best fitting Gaussian distribution. However, looking at Figure \ref{fig:gp_sp}, it is clear that neither distribution is close to Gaussian, and may not in fact have a clear mode to fit. 

Instead of approximating the posterior directly, I instead use the saddlepoint approximation.  The saddlepoint approximation for the distribution of random variables was originally proposed by \cite{danielsSaddlepointApproximationsStatistics1954}, with \cite{butlerSaddlepointApproximationsApplications2007} giving a thorough account of the basic method. Very roughly, the idea is to examine the cumulant generating function (CGF) for random variable $X$
\begin{equation}
    K(s) = \log \mathbb{E}[\exp(s X)] = \int_{\mathcal{X}} e^{s x} f(x) dx,
    \label{eq:cgf}
\end{equation}
where $f(x)$ is the probability distribution of $X$ and $\mathcal{X}$ is its domain of support. The existence of the CGF requires that there is some interval $a<0<b$ such that the above integral converges. Applying Laplace's approximation for this integral and rearranging terms, 
\begin{equation}
    \hat{f}(x) = \sqrt{\frac{1}{2\pi K''(\hat{s})}}\exp(K(\hat{s})-\hat{s} x),
\end{equation}
where $\hat{s}$ is the solution of $K'(s)=x$. $\hat{s}$ is a saddlepoint of the integrand in Equation \ref{eq:cgf}, hence the name ``saddlepoint approximation''. If the application requires it, $\hat{f}(x)$ then typically has to then be normalized to integrate to unity so that it is a true probability distribution, giving us
\begin{equation}
    \bar{f}(x) = \frac{\hat{f}(x)}{\int_{\mathcal{X}} \hat{f}(x)dx}.
\end{equation}
 If the application only requires the PDF up to proportionality (as is often the case), then the above normalization is not required, and the saddlepoint approximation requires no integration whatsoever. \cite{butlerSaddlepointApproximationsApplications2007} shows that this optimization problem is well posed and gives a unique real solution for $\hat{f}$, if $s$ is constrained to be inside the interval that contains $0$ for which $K(s)$ converges. Serendipitously, this low order method often provides extremely good approximations to the PDF, as the CGF $K$ contains the full information about the distribution of $X$. For sums of random variables (such as $||\nabla \tau||^2$), it is almost always easier to construct the CGF $K$ analytically rather than the PDF $f$, as $K_{X+Y}(s) = K_X(s)+K_Y(s)$, whereas $f_{X+Y}(x) = f_X(x) * f_Y(x)$ where $X$ and $Y$ are arbitrary random variables and $*$ is the convolution operator. Therefore, when using the saddlepoint approximation of to obtain the PDF, multiple potentially slowly converging convolution integrals are converted into a simple root-finding problem with a unique solution. Let us now apply this concept to deriving the PDFs of $||\nabla \tau||^2$ and $C_p$ from our closed form posteriors for phase delay derivatives $\nabla \tau$. To do this, my goal is to write the distribution of $||\nabla \tau||^2$ in a form for which I can determine the CGF $K_{||\nabla \tau||^2}$, and then use the saddlepoint approximation to obtain the posterior PDF $\hat{f}_{||\nabla \tau||^2}$, from which I can also obtain the posterior PDF $\hat{f}_{C_p}$ using a change-of-variables formula. 
 
 For simplicity, I approximate the posterior for a single point $x'$ given data $(\bf{X},\bf{y})$. I have shown that $\nabla \tau(x')|\bm{y} \sim N(\mu, \Sigma)$ for a $d$ dimensional mean vector $\mu$ and a $d\times d$ covariance matrix $\Sigma$. Therefore, 
\begin{equation}
\nabla \tau(x')|\bm{y} = Q\Lambda^{1/2}h + \mu,
\end{equation}
where $Q\Lambda Q^T = \Sigma$ is an eigenvalue decomposition of $\Sigma$ and $h$ is a $d$-dimensional standard normal variable $h \sim N(0,I)$. $Q$ contains the normalized eigenvectors as its columns and $\Lambda$ is a diagonal matrix of corresponding eigenvalues. Assuming that the phase delay measurements are taken in different locations, all of the terms in $\Lambda$ are positive as then $\Sigma$, as a non-degenerate covariance matrix, is positive definite. I can then write 
\begin{align}
    ||\nabla \tau(x')||^2 &= (Q\Lambda^{\frac{1}{2}}h + \mu)^T( Q\Lambda^{\frac{1}{2}}h + \mu) \nonumber\\
                      &= (Qh + \bar{\mu})^T\Lambda( Qh + \bar{\mu})\nonumber\\
                      &= (h+Q^T\bar{\mu})^TQ^T\Lambda Q(h+Q^T\bar{\mu})
                      \label{eq:quadform}
\end{align}
where $\bar{\mu} = \Lambda^{-\frac{1}{2}}\mu$. The eigenvalues collected in $\Lambda$ are labelled $\lambda_i$, with corresponding components of $\bar{\mu}$ labelled $\bar{\mu}_i$. The quadratic form in Equation \ref{eq:quadform} can be written as a sum over non-central chi-squared distributions \cite[]{imhofComputingDistributionQuadratic1961, butlerUniformSaddlepointApproximations2008}. The degree of freedom of each non-central chi-squared corresponds to the multiplicity of the eigenvalues of $\Sigma$, which will for our purposes always be distinct, giving
\begin{equation}
    ||\nabla \tau(x')||^2 = \sum_{i=1}^d \lambda_i \chi^2(1, \bar{\mu}_i^2).  
\end{equation}
Because of the summation property of the CGF, the CGF of $||\nabla \tau(x')||^2$ is then \cite[]{butlerUniformSaddlepointApproximations2008}
\begin{equation}
    K_{||\nabla \tau(x')||^2}(s) = \sum_{i=1}^d \left [-\frac{1}{2}\log (1-2s\lambda_i) + \frac{s \lambda_i \bar{\mu}_i^2}{1-2s\lambda_i} \right], 
    \label{eq:knt}
\end{equation}
and the derivatives are given by 
\begin{equation}
    K^{(j)}_{||\nabla \tau(x')||^2}(s) = 2^{j-1}(j-1)!\sum_{i=1}^d\lambda_i^j(1-2s\lambda_i)^{-j}\left(1+\frac{j\bar{\mu}_i^2}{1-2s\lambda_i}\right).
\end{equation}
The domain of convergence in which the root of $K'(s)=x$ is sought is the largest open interval containing zero for which $K_{||\nabla \tau(x')||^2}(s)$ is defined, which from looking at Equation \ref{eq:knt} is $s\in(-\infty, \frac{1}{2\lambda_{max}})$, where $\lambda_{max}$ is the largest eigenvalue of $\Sigma$.
Applying the saddlepoint approximation given the above $K$ gives us the saddlepoint distribution $\hat{f}_{||\nabla \tau(x')||^2}(x)$ for the squared slowness, which can be normalized to give
\begin{equation}
    \bar{f}_{||\nabla \tau(x')||^2}(x) =  \frac{\hat{f}_{||\nabla \tau(x')||^2}(x)}{\int_0^\infty \hat{f}_{||\nabla \tau(x')||^2}(x)}. 
\end{equation}
The transformation between squared slowness and phase velocity is given by $g(x) = \frac{1}{\sqrt{x}}$, which is a monotone decreasing function. The appropriate Jacobian transformation rule to obtain the approximate PDF of phase velocity is then \cite[]{kadanePrinciplesUncertainty2011}
\begin{equation}
    \bar{f}_{C_p}(x') = -\bar{f}_{||\nabla \tau(x')||^2}(g^{-1}(x)) \frac{d g^{-1}}{dx}(x) = \frac{2\bar{f}_{||\nabla \tau(x')||^2}\left(\frac{1}{x^2}\right)}{x^3}.
\end{equation}
The approximate distributions $\bar{f}_{||\nabla \tau(x')||^2}(x)$ and $\bar{f}_{C_p}(x')$ are plotted against a histogram of 1,000,000 draws of the squared slowness and phase velocity using the analytic derivatives in Figure \ref{fig:gp_sp}, showing that the saddlepoint approximations are a close fit. Higher order saddlepoint approximation terms and approximations for the cumulative distribution function (CDF) are collected in \cite{butlerSaddlepointApproximationsApplications2007}. 

The saddlepoint method can be further applied to the joint distribution function two points to derive the approximate spatial covariance \cite[]{al-naffouriDistributionIndefiniteQuadratic2016}. Because the underlying posterior distributions for the derivatives is given by a GP, the covariance completely describes the spatial behaviour of the velocity distribution, and so the ability to calculate the distribution for any two arbitrary points is sufficient to fully characterize the posterior. However, the resulting root-finding problem will be two-dimensional rather than one dimensional and is substantially more complicated than the forms derived here, so they are left for future work.

\section{Discussion}
\subsection{Implications for sample statistics}
Most eikonal tomography applications report per-station per-frequency error statistics by computing the standard error in the mean phase velocity over multiple sources. Studies typically appeal to the central limit theorem to justify the use of the sample standard error formula and sample mean for quantifying the data distribution. The reported standard errors are then used to weight data in further inversions --- a typical use case is to perform 1D Bayesian inversion beneath each station using the mean values and the reported error. Previous methods do not optimally smooth the phase delay regression that underlies eikonal tomography, potentially producing biased results, and do not produce uncertainty estimates for each source. However, uncertainties reported in studies using these methods are often extremely low, amounting to a few percent of the estimated phase velocity. 

\begin{figure}
    \centerline{\includegraphics{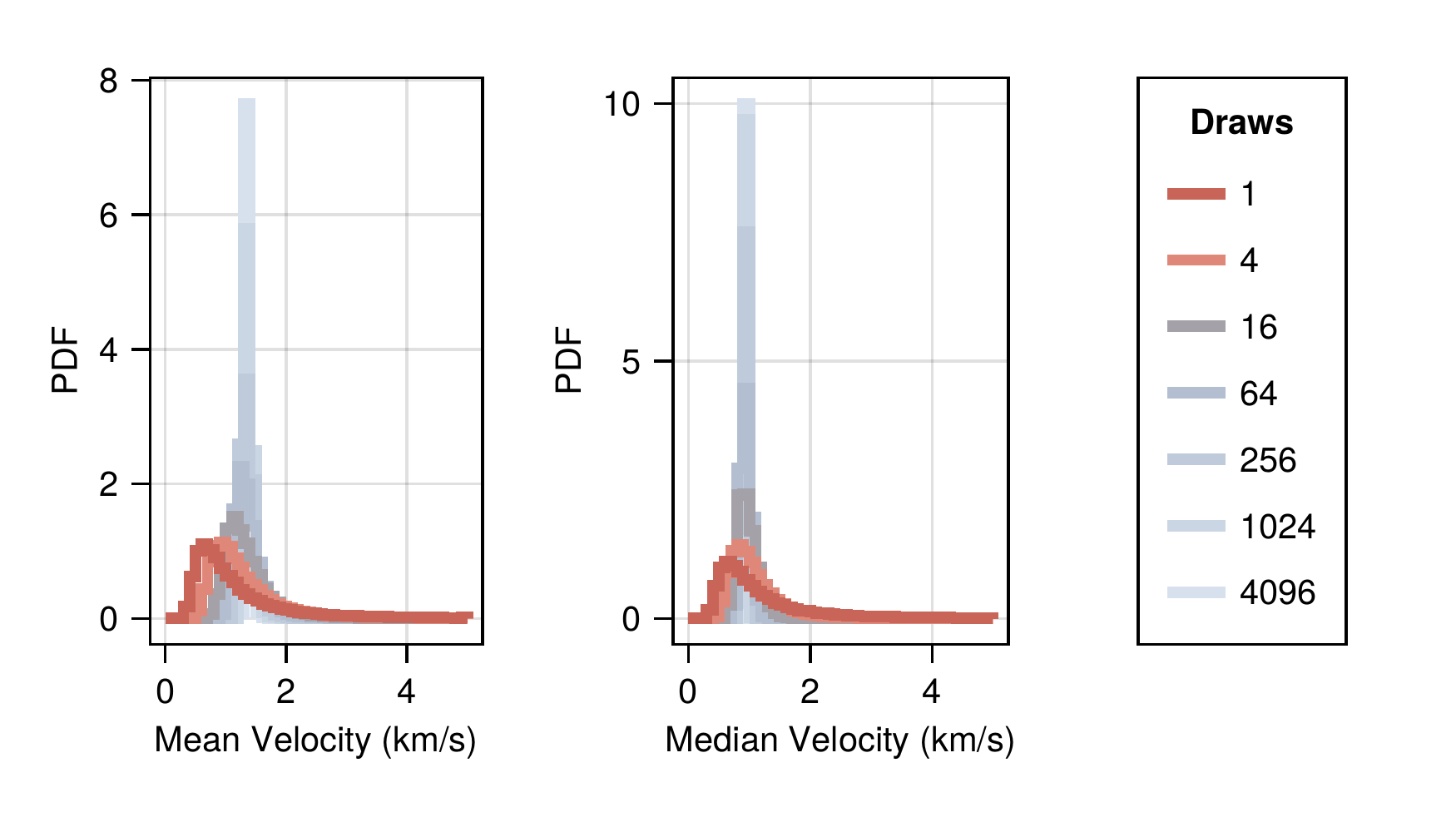}}
    \caption{Comparison of the distribution of sample means and sample medians for the phase velocity at (0.1875,0.3875). The mean or median is calculated by drawing $4^n$ samples for $n=0\dots6$. This process is repeated 100,000 times to obtain the distributions of sample means and medians. The sample mean converges to a normal distribution slowly.}
    \label{fig:gp_mean_vs_median}
\end{figure}

In our GP framework, Monte Carlo sampling can be used to directly estimate the distribution for sample statistics such as the mean over multiple sources. As a motivation, observe that both the empirical distribution for phase velocity and its saddlepoint approximation is heavy tailed in Figure \ref{fig:gp_sp}. This is a point relatively close to the edge, which can result in a distribution that is far from Gaussian. Taking this point, I  then draw $4^n$ samples of velocity for $n = 0\dots6$, calculate the sample mean and median, and then repeat 100,000 times to find the distribution in the sample statistics. Figure \ref{fig:gp_mean_vs_median} shows the results. The sample mean converges only slowly to a normal distribution, and is still broad even with 16 samples. In comparison, the sample median is well-behaved and converges quickly as the sample size increases. For both sample statistics, the distribution for small numbers of samples is unsurprisingly quite similar to the underlying velocity distribution, and is consequently heavy tailed --- this should be taking under consideration for applications such as fitting azimuthal anisotropy profiles to eikonal tomography results, where many azimuth bins near the edges of arrays will often have few contributing sources. 

\subsection{Future work}

In this study, I present the simplest possible implementation of a GP framework for eikonal tomography with analytic derivatives of phase delay. The flexibility of GP modelling offers several opportunities for future improvements that should result in more robust inversions. The first of these is that multi-frequency eikonal inversion is naturally handled by GP modelling by assuming a space-frequency covariance function. The most simple model would use a separable function $k((x,f), (x',f')) = k_x(x,x')k_f(f,f')$. A smooth frequency covariance $k_f(f,f')$ would reduce the impact of missing data in particular frequency bins, which can be an issue due to spectral holes in surface wave trains. 

Secondly, the squared-exponential kernel used in this study could be further improved to better represent the behaviour of true seismic wavefields; for instance, the problem could be recast in radial coordinates with a radial-azimuthal kernel as studied in \cite{padonouPolarGaussianProcesses2015}. Due to the natural cylindrical symmetry of wave propagation, this may allow us to reduce the uncertainty in the eikonal tomography results. In particular, this kernel choice would be appropriate in use cases such as ambient-noise tomography where the seismic source is inside the array, resulting in highly non-planar wavefronts. 

A third option would be to use the GP framework for smoothing the underlying full wavefield records before processing them for phase delay measurements or for other gradient based techniques such as wavefield gradiometry \cite[e.g.,][]{langstonSpatialGradientAnalysis2007, langstonWaveGradiometryTwo2007, deridderSurfaceScholteWave2015, deridderFullWaveField2018} or full Helmholtz tomography \cite[]{linHelmholtzSurfaceWave2011}. These applications would potentially require extending the GP derivative theory to higher order, but again noting that derivatives are linear, the resulting distributions for higher order spatial terms will also be GPs. The GP framework is especially well suited towards the inclusion of strain measurements in joint wavefield reconstruction \cite[e.g.,][]{muirWavefieldbasedEvaluationInstrument2021} as the appropriate covariance kernels can be calculated using the results in Equation \ref{eq:derivatives} --- an enticing prospect considering the proliferation of distributed acoustic sensing (DAS) strain sensors \cite[]{zhanDistributedAcousticSensing2020}. GP based techniques have also been used in geodesy to investigate transient strain rates \cite[e.g.,][]{hinesRevealingTransientStrain2018}, and the saddlepoint approximation techniques investigated here could offer a way to more accurate quantification of strain invariants arising from geodetic analysis. 

Finally, as the number of phase delay measurements increases across stations and frequency bins, the size of the data covariance matrix $\hat{K}$ increases. For $n$ measurements, the cost of inverting this matrix scales like $O(n^3)$, so very large collections of measurements pose a challenge for GP based inversion. Due to the popularity of GPs in machine learning research, there are a wide range of sparse GP approximations that produce almost identical results and still result in analytic derivatives once the sparsity structure is determined \cite[e.g.,][]{titsiasVariationalLearningInducing2009, lindgrenExplicitLinkGaussian2011, wilsonKernelInterpolationScalable2015}. Employing these methods would allow efficient upscaling of the methodology presented here to multi-frequency inversion of USArray-scale datasets. 

\subsection{Conclusions}

This study derives an analytic posterior distribution for phase delay derivatives, and then derives approximate posteriors for phase velocity using the saddlepoint approximation applied to the eikonal equation. The result is a fully Bayesian eikonal tomography that requires no MCMC sampling to characterize the posterior. As such, computations are easily implemented and highly efficient. Using the GP framework as a basis, I investigated two important effects that impact the interpretation of eikonal tomography results, namely the effect of the inclusion of data uncertainty on the expectation value of velocity and the behaviour of sample statistics, both of which suggest that the uncertainty in eikonal tomography results is greater than previously assessed. The GP framework presents a fully interpretable way forward to improve eikonal tomography in the future, with many opportunities for future work due to the flexible and robust nature of GP modelling. 

\section*{Data and code availability}
I have included the Pluto notebook used to generate the results in the submission. This notebook will be uploaded to Zenodo after acceptance so that the assigned DOI corresponds to the final version used for the publication.
    
\begin{acknowledgements}
    I acknowledge the support of the European Commission via Marie Skłodowska-Curie Actions Individual Fellowship award number 101027079. I also acknowledge Professor Ron Butler of Southern Methodist University for forwarding the technical report supporting \cite{butlerUniformSaddlepointApproximations2008}, and Professor Victor Tsai of Brown University for providing a pre-submission internal review that was very helpful. 
\end{acknowledgements}

\bibliography{references}

\end{document}